\documentclass{article} 
\usepackage{iclr2023_conference,times}

\usepackage{amsmath,amsfonts,bm}

\def\eqref#1{equation~\ref{#1}}

\def\1{\bm{1}}

\DeclareMathAlphabet{\mathsfit}{\encodingdefault}{\sfdefault}{m}{sl}
\SetMathAlphabet{\mathsfit}{bold}{\encodingdefault}{\sfdefault}{bx}{n}

\usepackage{hyperref}
\usepackage{url}
\usepackage{times}
\usepackage{latexsym}
\usepackage{tabularx}
\usepackage{booktabs}
\usepackage{multirow}
\usepackage{adjustbox}
\usepackage{caption}
\usepackage{subcaption}
\usepackage{tipa} 
\usepackage{enumitem}
\usepackage{amssymb}
\usepackage{amsthm}
\usepackage{pifont} 
\usepackage{array}
\usepackage{comment}
\usepackage{amsmath}
\usepackage{mathtools}

\title{Pretrained Vision Models for Predicting High-Risk Breast Cancer Stage}

\author{\normalsize Bonaventure F. P. Dossou $^{1}{^2}{^3}{^4}$, Yenoukoume S. K. Gbenou$^{5}$, Miglanche Ghomsi Nono$^{6}$\\\\
\footnotesize
$^1$Center for Intelligent Machines, McGill University, $^2$Mila Quebec AI Institute, $^3$Lelapa AI,\\$^4$Masakhane Research Foundation, $^5$Drexel University, $^5$University Of Maryland Baltimore County.
 \\}

\iclrfinalcopy 
\begin{document}

\maketitle

\begin{abstract}
Cancer is increasingly a global health issue. Seconding cardiovascular diseases, cancers are the second biggest cause of death in the world with millions of people succumbing to the disease every year. According to the World Health Organization (WHO) report, by the end of 2020, more than 7.8 million women have been diagnosed with breast cancer, making it the world’s most prevalent cancer. In this paper, using the Nightingale Open Science dataset of digital pathology (breast biopsy) images, we leverage the capabilities of pre-trained computer vision models for the breast cancer stage prediction task. While individual models achieve decent performances, we find out that the predictions of an ensemble model are more efficient, and offer a winning solution\footnote{https://www.nightingalescience.org/updates/hbc1-results}. We also provide analyses of the results and explore pathways for better interpretability and generalization. Our code is open-source at \url{https://github.com/bonaventuredossou/nightingale_winning_solution}
\end{abstract}
\section{Introduction}
Overall 2.3 million women have been diagnosed with breast cancer (BC) in 2020, and 685000 globally died from the disease. These statistics, coupled with the ones of the five last years, bring up to 7.8 million, the number of women alive who were diagnosed with BC by the end of 2020: more women have lost their lives due to BC than any other type of cancer, making it the most prevalent cancer in the world \footnote{Breast Cancer (WHO): \url{https://www.who.int/news-room/fact-sheets/detail/breast-cancer}}. The treatment of BC can be efficient when the disease is detected at a very early stage, and there are mainly five stages of BC (Stage 0, Stage 1, Stage 2, Stage 3, and Stage 4). One of the most popular ways of detecting BC is mammography, which is a detailed X-ray scanning (or screening) of the breast with Magnetic Resonance Imaging (MRI). Another approach is breast biopsy: a biopsy is a medical process during which samples of cell tissues are collected to be examined in the laboratory with a microscope. A biopsy helps to locate the presence, cause, or extent of the disease.

Thanks to recent advances in Artificial Intelligence (AI), especially Deep Learning (DL), there has been a rise in research efforts to leverage the potential of DL-based systems to help in breast cancer detection. \cite{bc1} provided a narrative review of AI in mammographic screening of breast cancer risk: the recent initiatives made use of pretrained Computer Vision (CV) models like ResNets \cite{resnet}, DenseNets \cite{densenet}, AlexNet \cite{alexnet}, U-Net \cite{unet}, cGAN \cite{cgan}, \cite{googlelenet}, MobileNet \cite{mobilenet}, Inception \cite{inception}, and RetinaNet \cite{retinanet}. All of these methods were applied to Full-Field Digital Mammography (FFDM) in which X-rays mammograms are converted into electrical signals, in a binary classification (detecting BC or not BC) setting.

In this work, after briefly describing the Nightingale Open Science Dataset of Digital Pathology (NOSDDP) \cite{dataset, dataset1}, we use slides from each biopsy digital image and pretrained CV models, to predict the stage of cancer of a patient (see Figure \ref{task_description}).
\section{Datasets}
\begin{figure*}[!ht]
    \centering
\includegraphics[width=0.8\linewidth]{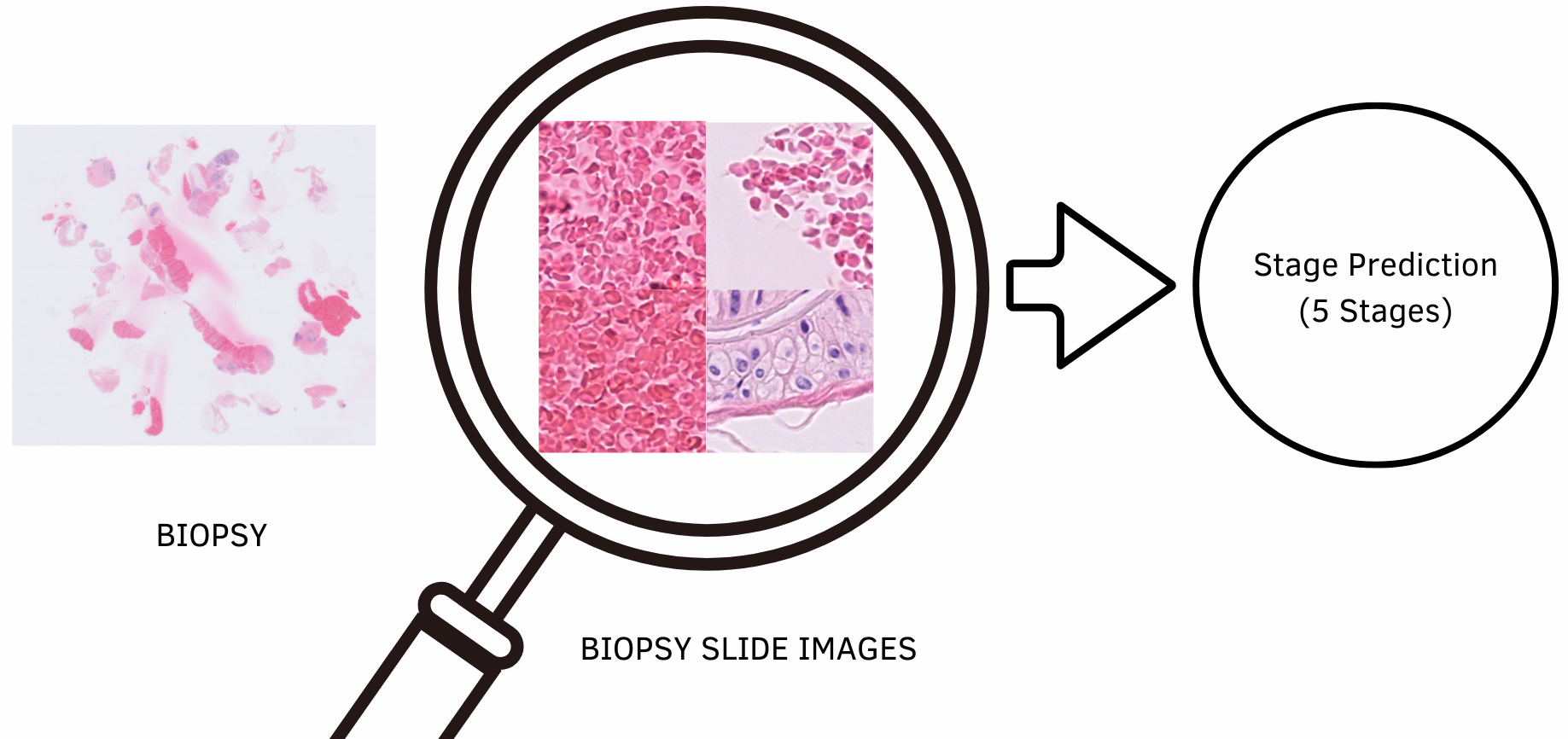}
    \caption{Description of the task: a biopsy generates slide images, that can be considered as different patches (parts) of the original biopsy. The cancer stage for the entire biopsy is the average of the individual stage prediction of each slide image.}
    \label{task_description}
\end{figure*}

\begin{figure*}[!ht]
    \centering
\includegraphics[width=0.8\linewidth]{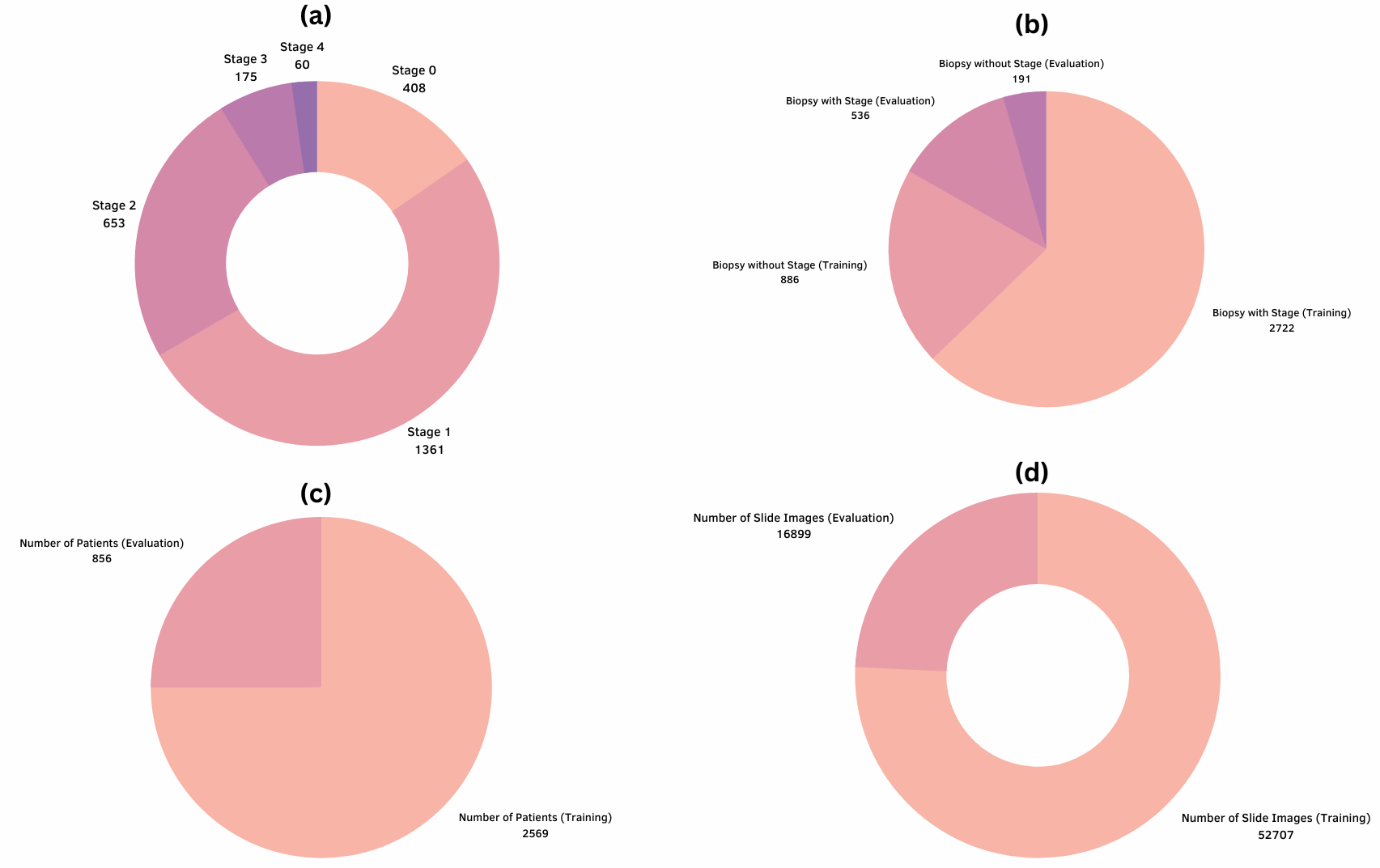}
    \caption{Statistics of the dataset: (a): Distribution of biopsies per cancer stage, (b): Distribution of labeled biopsies and non-labeled biopsies across the different dataset splits, (c): Number of patients used for each dataset split, and (d): Number of slide images corresponding to biopsies collected across the different dataset splits.}
    \label{dataset_distribution}
\end{figure*}

Existing works have demonstrated the ability of DL-based systems to predict the type of cancer based on X-ray scans and mammograms. DL-based algorithms also focus on features that sometimes get overlooked or neglected by specialists (for instance the nature of the non-cancerous tissue surrounding the tumor) \cite{doi:10.1126/scitranslmed.3002564}. However, there are currently very few to no datasets, that link biopsy images to the cancer stage of the patients, and the respective outcomes: the NOSDDP is an attempt to bridge that gap. The NOSDDP contains 72400 biopsy slide images, from 4335 breast biopsies of 3425 unique patients, observed from 2014 to 2020. These slide images are linked to cancer stages and information about the mortality of patients (see Figure \ref{task_description}). Figure \ref{dataset_distribution} presents the statics across the dataset, and its different splits (train and evaluation). The main takeaway is that most patients across the dataset are respectively in BC Stages 1, 2, and 0. Stages 1, and 0 are considered \textit{early}, while Stage 2 can already be the characteristic of a \textit{very advanced} BC. It is also important to notice that there are biopsies without stages (i.e. biopsies that have not been labeled with a specific cancer stage). In the next section, after describing our experimental setups which are: (1) the preprocessing of slide images, and (2) the pretrained CV models we use, we present the results and analyses from the different pretrained CV models.

\section{Experiments and Results}
We make complete use of available data by assigning to the unlabeled biopsies, Cancer Stage 1 (the most frequent cancer stage). We then downsampled the high-resolution slide images to (224, 224), as supported by many existing pretrained CV models. The original training set has been split into two subsets (with 80:20 ratio): $D_{Train}$ and $D_{Eval}$, while the initial evaluation set is used as a test dataset, we denote it $D_{Test}$. Image samples from $D_{Train}$ have been augmented using randomized cropping and horizontal flipping, then normalized. The image samples from $D_{Eval}$ and $D_{Test}$ have only been normalized.

We finetuned 10 pretrained CV models: Resnet (18, 50, 152), EfficientNet \cite{efficientnet}, ConvNext \cite{convnext} (M), WideResNet \cite{wideresnet} (101), VGG \cite{vgg}, ResNext \cite{resnext} (101), RegNet \cite{regnet} (X32GF), Swin Transformer \cite{swin} (B), and MaxVit \cite{maxvit}. These models are all high-performing computer vision models, and their respective versions (or size) were chosen to cope with the memory available. For each model, we try various learning rates, with the \textit{AdamW} optimizer. Each model was trained with a batch size of 32, and for 50 epochs: those were the efficient values for those hyperparameters. The predicted cancer stage $PCS$ for a biopsy image $B$ is the average of the predicted cancer stage for each subsequent slide image $S$:
\[PCS(B) = \frac{1}{|B|}\sum_{i} PCS(S_{i})\] where $|B|$ is the number of slides for the biopsy $B$, and $PCS(S_{i}) \in \{0, 1, 2, 3, 4\}$: this implies that the prediction for a biopsy image is a continuous value $\in [0, 4]$. Therefore, to measure the closeness to the true label which $\in \{0, 1, 2, 3, 4\}$, we use the Mean Square Error (MSE) metric:
\[MSE = \frac{1}{n} \sum_{j=1}^{n} (\hat{y}_{j}–y_{j})^{2}\] where the $actualstage \in \{0, 1, 2, 3, 4\}$. The different performances are summarized in Table \ref{training_results}.

\begin{table*}[!ht]
\footnotesize
 \begin{center}
 \resizebox{\textwidth}{!}{
   \begin{tabular}{cccccccccccc}
    \toprule
    \textbf{Learning Rate}&\textbf{Resnet 18}&\textbf{Resnet 50}&\textbf{Resnet 152}&\textbf{EfficientNet (M)}&\textbf{ConvNext (Base)}&\textbf{Wide Resnet 101}&\textbf{VGG}&\textbf{RegNet}&\textbf{SwinT (B)}&\textbf{MaxVit} \\ 
    \midrule
    1e-4 & 1.009611 & \textbf{0.970504} & 1.005862& 0.925831 & 1.009943 & \textbf{0.931212} & \textbf{0.939045} & 0.992109&0.898370&\textbf{0.855656}\\
    1e-5 & \textbf{1.001620} & 1.008508 & \textbf{1.001902} & 0.987911 & \textbf{0.957443} & 1.012101& 0.994687& \textbf{0.988756} &\textbf{0.897878}&0.893357\\
    4e-4 & 1.020936 & 0.986704 & 1.007281 & \textbf{0.829745} & 1.110100 & 1.033937 & 1.107765&1.000450& 1.231371&1.098370\\
    \bottomrule
  \end{tabular}}
    \caption{MSE of each pretrained CV model. Arranged per column (model), \textbf{the bold numbers represent the best performance of each model, with respect to learning rates}.} \label{training_results}
  \end{center}
\end{table*}

\begin{figure*}[!ht]
    \centering
\includegraphics[width=0.8\linewidth]{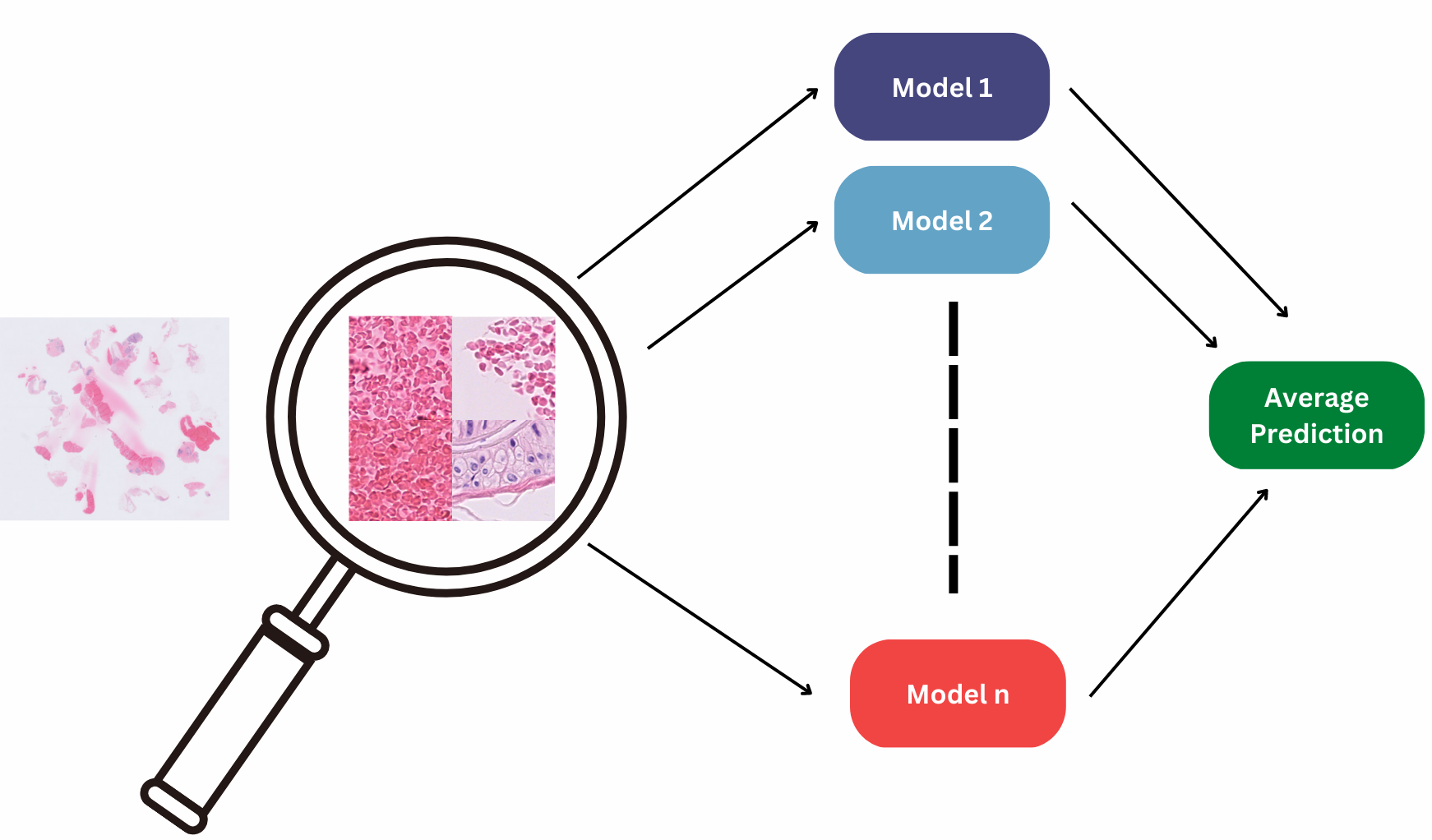}
    \caption{Deep Ensemble of each pretrained CV model: The average prediction for a biopsy is the average prediction of each model.}
    \label{deep_ensembles}
\end{figure*}

Looking at individual performances, we can see that EfficientNet performs better than other models. We believe this is due to their architecture, which combines families (like an ensemble) of individual baseline Neural Networks (NNs). Each NN uses a mobile inverted bottleneck convolution architecture, to optimize their respective accuracy and efficiency via neural architecture search \cite{efficientnet}\footnote{EfficientNet: Improving Accuracy and Efficiency through AutoML and Model Scaling (GoogleAI Blog): \url{https://ai.googleblog.com/2019/05/efficientnet-improving-accuracy-and.html}}. Following the intuition of EfficientNet, we also tried to create a Deep Ensemble ($E$) of individual pretrained CV models (see Figure \ref{deep_ensembles}). 
Therefore, the predicted cancer stage of the ensemble model $E$, for a biopsy $B$ is \[PCS(B)_{E} = \frac{1}{|E|} \sum_{k=1}^{|E|} PCS(B)_{k}\] 

where $PCS(B)_{k}$ is the $PCS(B)$ of the $k$-th model, and $|E|$ is the size of $E$ i.e. the number of model composing $E$: the predicted cancer stage of the ensemble model $E$, for a biopsy $B$ is hence the average of the predictions of each model of the ensemble $E$ for that biopsy $B$.

We explore two strategies: (a) deep ensemble with all models, and (b) deep ensemble solely with the models which have an MSE lower than 1. We find out that in setup (a), the MSE is 0.632767 while in setup (b) the MSE is \textbf{0.5543481} (which offered a winning solution)\footnote{Nightingale Predicting High-Risk Breast Cancer Contest: \url{https://www.nightingalescience.org/updates/hbc1-results}}. This shows that: (*) Deep Ensembles are better than individual models. This makes sense because in real-life, obtaining and aggregating the predictions of many experts (doctors) is better than relying on the opinion of a sole doctor. Moreover, each model learns different representations and features, which reduces the bias toward specific classes; and (**) the lower the losses of individual models, the lower the MSE of the Deep Ensemble. 

\section{Looking Beyond Predictions: Exploring Causal Inference for a better Interpretability, Performance, and Generalization}
Many high-performing models have failed when deployed in real-world settings, due to distribution shifts. Consequently, many efforts are put into interpretability and enhancement of robustness. Causal Inference (CI) approaches could benefit medical imaging problems (the BC can be considered as a type of medical imaging problem), as it would help understand the causal relationships between features (variables) and how they could potentially influence models predictions. CI is little to not explored in the Medical Imaging field, but its applications could be very beneficial to it, and even to healthcare and precision medicine \cite{causalinference1, causalinference2, causalinference3, causalinference4, causalinference5}. Understanding \textit{causal-effect} relationships and their influences on the predictions, will help increase robustness, and reduce bias while providing more interpretable predictions to medical practitioners (e.g. doctors), allowing more \textit{trust}.
\section{Conclusion}
In this work, we explored the applicability of pretrained computer vision models, in the task of predicting high-risk breast cancer. We observed that Deep Ensemble models offer better performances, than single models. Our approach produced a winning solution for Nightingale's Breast Cancer context, and we open-source our code at \url{https://github.com/bonaventuredossou/nightingale_winning_solution}. As future work, we are looking forward to exploring the practicability of causal inference to medical imaging challenges, but also how uncertainty estimation could be useful in the medical imaging context particularly.

\bibliography{iclr2023_conference}
\bibliographystyle{iclr2023_conference}

\appendix

\end{document}